\documentclass[useAMS,usenatbib]{mnras}
\def\apj{ApJ}
\def\apjs{ApJ}
\def\mnras{MNRAS}
\def\aap{A\&A}
\def\apjl{ApJ}

\def\nat{Nature}

\def\araa{Annual Reviews}

\usepackage{fancyhdr}
\usepackage{graphicx,latexsym}
\usepackage{color}
\usepackage{xcolor}
\usepackage{mathtools}
\usepackage[margins]{trackchanges}
\usepackage{multirow}
\usepackage{subcaption}
\usepackage{float}
\usepackage{ctable}
\usepackage[hyphenbreaks]{breakurl}
\usepackage{array}
\usepackage{color,soul}
\usepackage[section]{placeins}
\usepackage{bigints}
\usepackage{hyperref}
\usepackage{bm}
\usepackage{amsmath,amssymb}
\usepackage{pdflscape}
\usepackage{scrhack}
\usepackage{tabularx}
\usepackage{booktabs}
\usepackage{siunitx}
\usepackage{colortbl}


\newcommand{\GG}[1]{}

\addeditor{LdB}

\usepackage{times}

\newcolumntype{M}{>{\centering\arraybackslash}m{\dimexpr.35\linewidth-2\tabcolsep}}
\newcolumntype{N}{>{\centering\arraybackslash}m{\dimexpr.16\linewidth-2\tabcolsep}}
\newcolumntype{Z}{>{\centering\arraybackslash}m{\dimexpr.23\linewidth-2\tabcolsep}}
\newcolumntype{B}{>{\centering\arraybackslash}m{\dimexpr.20\linewidth-2\tabcolsep}}

\title[Mass/spin estimates for 4U1608-52 through the RPM]{Mass and spin measurements for the neutron star 4U1608-52 through the relativistic precession model}
\author[du Buisson et al.]{L. du Buisson$^{1}$\thanks{E-mail:
lise.dubuisson@chch.ox.ac.uk}, S. Motta$^{1,2}$ and R. Fender$^{1,3}$\\
$^{1}$Department of Physics, University of Oxford, Denys Wilkinson Building, Keble Road, Oxford, OX1 3RH, UK.\\
$^{2}$ESAC, European Space Astronomy Centre, Villanueva de la Ca\~nada, E-28692 Madrid, Spain.\\
$^{3}$Department of Astronomy, University of Cape Town, South Africa.}

\begin{document}

\raggedbottom

\date{Accepted 2019 April 22. Received 2019 April 16; in original form 2019 January 21}

\pagerange{\pageref{firstpage}--\pageref{lastpage}} \pubyear{2018}

\maketitle

\label{firstpage}

\begin{abstract}
We present a systematic analysis of the complete set of observations of the neutron star low-mass X-ray binary 4U1608-52 obtained by the \textit{Rossi X-ray Timing Explorer's Proportional Counter Array}. We study the spectral and fast-time variability properties of the source in order to determine the mass and spin of the neutron star via the relativistic precession model, and find 24 observations containing usable sets of the necessary three quasi-periodic oscillations (triplets) with which to accomplish this task, along with a further 7 observations containing two of the three quasi-periodic oscillations each. We calculate the spin and mass of the source for each of the triplets, and find that they give physically realistic estimates clustering in the spin range $0.19 < a < 0.35$ and mass range $2.15 < M/\textrm{M}_\odot < 2.6$. Neutron stars present environments for studying matter under the most extreme conditions of pressure and density; as their equation of state is not yet known, accurate measurements of their mass and spin will eventually allow for the discrimination between various models. We discuss the implications of our findings in the context of equation of state predictions, physically allowed spin ranges, emission proximity to the innermost stable circular orbit and possible model inaccuracies.
\end{abstract}






\section{Introduction}
\label{sec:intro}
Low-mass X-ray binaries (LMXBs) are systems hosting either a black hole (BH) or a neutron star (NS) that accretes mass from a companion star via an accretion disk radiating in the X-ray \citep{Shakura1973}. While NS systems are mainly persistent systems, BH systems tend to be transients \citep{Munoz2014, Motta2017}. NS LMXBs have been historically classified as either {\it Atolls} or {\it Z-sources}, based on the shapes they trace in their Colour-Colour diagrams (CCDs, \citealt{Hasinger1989}). \cite{Munoz2014} showed that NS LMXBs can be described in the same state/transition scheme typically used for BH systems, and that rms-intensity diagrams (RIDs, \citealt{Munoz2011}) and hardness-intensity diagrams (HIDs, \citealt{Homan2001}) can be used in the same way when determining timing and spectral properties of both of these classes of systems. The different states of NS and BH LMXBs are defined based on both their X-ray spectral properties and the inspection of features in the X-ray Fourier power density spectra (PDS) of their observations, the latter used for studying changes in fast time variability. The PDS features of most interest to us are narrow peaks called quasi-periodic oscillations (QPOs).

QPOs were first discovered in cataclysmic variables by \mbox{\cite{patterson1977},} making use of the 76 cm and 2.1 m telescopes of the McDonald Observatory\footnote{McDonald Observatory: \burl{https://mcdonaldobservatory.org/}}. They have since been discovered in the light curves of NS and BH LMXBs by the European X-ray Observatory Satellite (EXOSAT\footnote{EXOSAT: \burl{https://www.cosmos.esa.int/web/exosat}}) and the Ginga satellite\footnote{Ginga: \burl{https://heasarc.gsfc.nasa.gov/docs/ginga/ginga.html}} (see e.g. \citealt{vanderklis1985, middleditch1986, Motch1983, Miyamoto1991}), and have also been found in ultra-luminous X-ray sources (e.g. \citealt{strohmayerULX2003, bachetti2014}) and in Active Galactic Nuclei (AGN, e.g. \citealt{gierlinski2008, middleton2010}). Despite being known for decades now, QPOs remain poorly understood, and their physical origin is still largely debated, even though it is commonly accepted that they must originate in the innermost regions of the accretion flow. The centroid frequencies of QPOs can be measured with high accuracy, giving us the opportunity to probe the distorted spacetime in the strong gravitational field regime close to compact objects \citep{bookvanderklis}.


Throughout its years of operation, the Rossi X-ray Timing Explorer (RXTE) \footnote{RXTE: \burl{https://heasarc.gsfc.nasa.gov/docs/xte/}} yielded an in-depth phenomenological \mbox{knowledge} regarding QPO observational properties. In both NS and BH systems QPOs have been divided into high- and low-frequency QPOs. Low-frequency QPOs (LF QPOs) in NS LMXBs have centroid frequencies ranging between $\sim 0.1$ Hz and $\sim 60$ Hz. For Z sources, LF QPOs are divided into normal branch oscillations (NBOs, \citealt{middleditch1986}), horizontal branch oscillations (HBOs, \citealt{vanderklis1985}) and flaring branch oscillations (FBOs, \citealt{Klis1989}). Although LF QPO classification is less obvious for Atoll sources \citep{Straaten2001}, they have also been divided into groups \citep{DiSalvo2003}, further studied and named by \cite{Motta2017}: HBO-like QPOs in the range from mHz to $\sim 40$ Hz, and FBO-like QPOs. NS high-frequency QPOs (HF QPOs, \citealt{Stroh2001, belloni2012}) are called kHz QPOs (divided into upper and lower kHz QPOs) with centroid frequencies in the range $\sim 400$ Hz to above $1$ kHz - they often occur in a pair, are regularly observed and usually have very high amplitudes \citep{Motta2017, vanderklis1996}. In BH LMXBs, LF QPOs range from $\sim 0.1$ Hz to $\sim 30$ Hz and are classified as either Type-A, -B or C QPOs (see \citealt{Wijnands1999a, Casella2005, Motta2012}), while QPOs with centre frequencies $\gtrsim 100$ Hz and up to $\sim 500$ Hz are referred to as upper or lower HF QPOs. An association between Type A, -B and -C QPOs in BHs with FBOs, NBOs and HBOs in NSs, respectively, has been proposed by \mbox{\cite{Casella2005}.}

There are different groups of suggested QPO mechanisms in the literature. One group incorporates wave modes of the accretion flow (e.g. \citealt{tagger1999, titar1999, wagoner2001, cabanac2010}), while another uses the general-relativistic hydrodynamical simulations of thick tori orbiting compact objects and explains QPOs through the oscillation modes of thick disks (e.g. \citealt{comment3_1, comment3_2, comment3_3, comment3_4, comment3_5}). Lastly, there are those models that are associated with relativistic effects involving the misalignment of the compact object's spin and the accretion flow (e.g. \citealt{Stella1998, lamb2001, Abram2001, fragile2005, schnitt2006, homan2006, ingramdone2011}). The latter models are based on the fact that General Relativity (GR) predicts that the presence of a spinning mass will affect the motion of matter around it. In these systems nodal, or Lense-Thirring, precession occurs when the spin axis of the compact object is misaligned with the orbital axis, so that there is a precession of the orbit's spin axis around the compact object's spin axis with a frequency $\nu_\textrm{nod} = \nu_\phi - \nu_\theta$ (with orbital frequency $\nu_\phi$ and vertical frequency $\nu_\theta$). Periastron precession, on the other hand, is the precession of the elliptical orbit's semi-major axis with a frequency $\nu_\textrm{per} = \nu_\phi - \nu_\textrm{r}$ (with radial epicyclic frequency $\nu_\textrm{r}$). The relativistic precession model (RPM, see section \ref{sec:rpm}) associates $\nu_\textrm{nod}$, $\nu_\textrm{per}$ and $\nu_\phi$ each with a respective QPO - observed simultaneously, this set of three QPOs are known as a 'triplet' and can be used to test the RPM.


4U1608-52 is a fairly bright transient NS Atoll source that has been sparsely monitored over a 16-year period with RXTE. It shows a very rich fast time-variability phenomenology: it contains twin kHz QPOs, LF QPOs, likely Hecto-Hz QPOs, Type-I X-ray bursts and burst oscillations. Thanks to these burst oscillations, the spin of 4U1608-52 has been measured to be 619 Hz, making it one of the most rapidly rotating accreting NSs \citep{gallo}.

In this study we characterise 4U1608-52 in terms of its fast time-variability and spectral behaviour using all available RXTE Proportional Counter Array (PCA) data on the source. We find altogether 31 QPO triplets with which to test the RPM - this is the first time that more than one triplet from the same BH or NS system is used to test the RPM quantitatively, and it therefore enables us to more accurately test the reliability of the model in determining the mass and spin of neutron stars in LMXBs.

NSs are natural laboratories for studying some of the most extreme conditions of pressure and density found in the Universe, and their equation of state (EoS) is as of yet still unknown \citep{WhyDisInNature}. The mass and spin are essential parameters for determining and constraining the Eos of NSs, and the RPM supplies us with a way in which to obtain independent measurements of both these quantities, thereby playing a potentially important role in uncovering the extreme physics governing NS interiors. 

This paper is structured as follows. Section \ref{sec:rpm} outlines the analytic solution to the RPM while section \ref{sec:data} details our timing and spectral analysis, selection of observations, the power spectral fitting and the classification of our observed QPO sets. \mbox{Section \ref{sec:results}} states our results, section \ref{sec:disc} discusses them and section \ref{sec:summ} summarises and concludes the study.


\section{The relativistic precession model}
\label{sec:rpm}
The RPM was proposed by \cite{Stella1998} and associates Lense-Thirring precession with the Type-C QPO (or HBO or HBO-like QPO in NS systems), periastron precession with the lower \mbox{HF QPO} (lower kHz QPO in NS systems) and the orbital frequency with the upper HF QPO (upper kHz QPO in NS systems). With all three signals emitted from the same characteristic \mbox{radius $r$}, the inward motion of the accretion inner truncation radius can be used to explain the co-evolution of these QPOs towards higher frequencies \citep{rpm2}. Although results with varying levels of success have been achieved when applying the model to NS systems (\citealt{stella1999, done1, terzan5}), the RPM works well for BHs. Examples of this are the papers by \cite{m2014} and \cite{Motta2014b}, where results are consistent with previous findings in the literature.

An argument against the test particle RPM is that it is expected that any local disc feature (the emission of which is responsible for the appearance of QPOs) will be sheared out by differential rotation (see e.g. the discussion in \citealt{comment3_6}, where they compare their model with the RPM). The rigid precession model (\citealt{Ingram2009}) tackles this problem by making the global, rigid precession of a toroidal structure responsible for the existence of the Type-C QPO, instead of the precession of a single test particle. For this, the ``truncated disk model" geometry is assumed: a cool ($T \sim 1$ keV), optically thick, geometrically thin accretion disk that is truncated at some radius, filled with a hot ($T \sim 100$ keV), geometrically thick inner accretion flow, where Type-C QPOs arise due to the Lense-Thirring precession (due to the effect of frame-dragging) of the hot inner flow (\citealt{Ingram2009, Done2012, lodato2013, rigidpmMotta}). The RPM asymptotically tends to the rigid precession model as the emission radius $r$ approaches the innermost stable circular orbit (ISCO) [see Figure 2 in \cite{rigidpmMotta}].

Initially it was thought that the equations yielded by the RPM had no analytical solution, making them very computationally intensive, but \cite{IngramRPM2014} found an analytical solution for the case where three simultaneous QPOs (a triplet) are detected - these can be used to determine the mass $M$ and spin $a$ of the compact object, as well as the emission radius $r$. Their solution is discussed in the remainder of this section.

In a system where a test mass orbits a spinning BH in a plane that is slightly perturbed from the equatorial, the orbital ($\nu_{\phi}$), periastron ($\nu_{\textrm{per}}$) and nodal precession ($\nu_{\textrm{nod}}$) frequencies can be shown to be

\begin{equation}
\nu_{\phi} = \pm \frac{\beta}{M} \frac{1}{r^{3/2} \pm a}
\label{eq:nuphi}
\end{equation}

\begin{equation}
\nu_{\textrm{per}} = \nu_{\phi} \left( 1 - \sqrt{1 - \frac{6}{r} \pm \frac{8a}{r^{3/2}} - \frac{3a^2}{r^2}} \right)
\label{eq:nuper}
\end{equation}

\begin{equation}
\nu_{\textrm{nod}} = \nu_{\phi} \left( 1 - \sqrt{1 \mp \frac{4a}{r^{3/2}} + \frac{3a^2}{r^2}} \right) \hspace{1mm} ,
\label{eq:nunod}
\end{equation}

\noindent using the Kerr metric (\citealt{bardeen, merloni}). Above, $r$ is the radius of the test particle's orbit in units of \mbox{$R_g = GMM_{\odot}/c^2$}, where $G$ is the gravitational constant, $c$ is the speed of light and $M$ is the BH mass in solar masses. The dimensionless spin parameter of the compact object is defined as $a = (cJ) / (GM^2)$, where $J$ is the angular momentum of the object expressed as $J = I\omega$, $I$ being the moment of inertia and $\omega$ being the angular frequency expressed as $\omega = 2\pi\nu$, with $\nu$ being the spin frequency of the compact object. Lastly, \mbox{$\beta = c^3 / (2\pi G M_{\odot}) = 3.237 \times 10^4$ Hz.} In all equations, the upper sign refers to prograde spin and the lower sign denotes retrograde spin, where prograde spin is defined as orbital motion in the direction of BH spin. As no stable orbit can exist inside the ISCO, the condition $r > r_{\textrm{ISCO}}$ can further be set, where $r_{\textrm{ISCO}}$ depends on the spin monotonically \mbox{(\citealt{bardeen, m2014}).} In the case where all three of the triplet QPOs are present in an observation, and under the simplest assumption that they all arise from the same radius $r$, we can use these solutions to determine $a$, $M$ and $r$, where

\begin{equation}
r = \frac{2}{3} \frac{6 - \Delta - 5\Gamma + 2\sqrt{2(\Delta-\Gamma)(3-\Delta-2\Gamma)}}{(\Delta+\Gamma-2)^2}
\label{eq:r}
\end{equation}

\begin{equation}
a = \pm \frac{r^{3/2}}{4} \left( \Delta + \Gamma - 2 + \frac{6}{r} \right)  \hspace{1mm} ,
\label{eq:a}
\end{equation}

\noindent and where $\Gamma$ and $\Delta$ are given by

\begin{equation}
\Gamma \equiv \left( 1 - \frac{\nu_{\textrm{per}}}{\nu_\phi} \right)^2 = 1 - \frac{6}{r} \pm \frac{8a}{r^{3/2}} - \frac{3a^2}{r^2}
\label{eq:Gamma}
\end{equation}

\begin{equation}
\Delta \equiv \left( 1 - \frac{\nu_{\textrm{nod}}}{\nu_\phi} \right)^2 = 1 \mp \frac{4a}{r^{3/2}} + \frac{3a^2}{r^2} \hspace{1mm} .
\label{eq:Delta}
\end{equation}

\noindent After determining $r$ and $a$ in this way, the spin can then be found using Eq. \ref{eq:nuphi}. Due to the non-linearity of the equations in $a$, $M$ and $r$, error estimates are obtained using a Monte Carlo simulation and following the procedure provided in \cite{IngramRPM2014}.

The simultaneous occurrence of the three QPOs is rare. However, it is also possible to use two simultaneous QPOs and an independent mass measurement to estimate the spin, as shown by \cite{IngramRPM2014}. If no such mass measurement is available, \cite{IngramRPM2014} show how limits can be placed on the system using only two out of the three essential QPOs.

\section{Observations and Data Analysis}
\label{sec:data}
We analysed all publicly available archival RXTE PCA data on the NS LMXB 4U1608-52, observed over a time period from March 1996 to December 2011. We removed time segments containing \mbox{Type-I} X-ray bursts from observations and excluded all observations with source count rates below 10 cts/s/PCU\footnote{PCU: proportional counter unit} to ensure an adequately high signal-to-noise ratio (SNR). Along with the cutting of certain long observations into shorter independent segments (see \mbox{section \ref{subsec:selection}),} this left us with a total of 1030 usable observations.


\subsection{Timing analysis}
\label{subsec:timing}
For each observation we considered Event, Binned, Single Bit and Good Xenon PCA data modes \citep{jahoda1996, Bradt1993} and calculated the PDS using custom software under IDL\footnote{GHATS: \hspace{1mm} \burl{http://www.brera.inaf.it/utenti/belloni/GHATS_Package/Home.html}} in the energy band $2-115$ keV (absolute PCA channels $0$ to $249$). We used a time resolution of $1/8192$ s ($\sim 122 \hspace{1mm} \mu s$) where possible, and divided each observation into intervals of 16 seconds. Averaging the Leahy-normalised PDS (without subtracting the Poissonian noise, which we fit together with the source PDS) for each observation, these choices resulted in PDS with a Nyquist frequency of $4096$ Hz and a frequency resolution of $0.0625$ Hz. We then convert the PDS to rms-normalised PDS following \cite{belloni1990}, and measure the fractional root mean square (rms) of the data in the $0.1 - 64$ Hz range \citep{vanderklis1989}. 


\subsection{Spectral analysis}
\label{subsec:spec}
The count rates necessary for the computation of the HID were obtained using the Standard 2 energy spectra extracted from PCU \mbox{unit 2} that covers the $2-60$ keV energy band across 129 channels. For each observation, the source count rate was determined using the $2-16$ keV energy band with the hardness measure defined as the ratio of counts in two energy bands $A$ and $B$ as $H_{\textrm{HID}} = A/B$, where $A$ stretches between $6-10$ keV and $B$ stretches between $4-6$ keV. The Standard 2 channels corresponding to these energy ranges shift slightly depending on the corresponding RXTE gain epoch\footnote{RXTE gain epochs: \hspace{1mm} \burl{https://heasarc.gsfc.nasa.gov/docs/xte/e-c_table.html}} and were adjusted for accordingly.

\subsection{Selection of observations and power spectral fitting}
\label{subsec:selection}
In order to find the QPOs present in the dataset, we preselected the PDS of observations that by visual inspection contained these features. Due to the movement of QPOs in frequency as time progresses, it often happens that the resulting averaged PDS of a long observation contains significantly broadened features, where a `long observation' is here defined as being of the order of roughly $12000$ seconds or more in length. In such cases the observation was cut into shorter consecutive independent segments, and an averaged PDS for each of these resulting shorter observations was then calculated and used for the fitting and classification of QPOs. The main focus of this work is the testing of the RPM using the nodal, periastron precession and orbital frequencies of the system. It is therefore specifically important for us to identify PDS of observations which simultaneously show the occurrence of the associated three features: an HBO-like QPO, a lower kHz QPO and an upper kHz QPO.

The narrow and broad features of each power spectrum were fit with Lorentzians and a power-law component by means of the XSPEC\footnote{XSPEC: \hspace{1mm} \burl{https://heasarc.gsfc.nasa.gov/xanadu/xspec/}} package; Lorentzians have their peak at the centroid frequency $v_c$ of a feature. We excluded all non-significant features from the analysis: for very low frequencies where red noise or flat-top noise were present, this meant excluding features below a significance\footnote{Significance is calculated as the integral of the power of the Lorentzian used for the fitting of the feature divided by the negative $1\sigma$ error on this integral.} of $2\sigma$; at all other frequencies features had to be significant at or above $3\sigma$. For a feature to be classified as a QPO its significance had to be at or above $3\sigma$ and its quality factor $Q$ (a measure of the coherence of the Lorentzian) had to be $Q \geq 2$, where $Q = v_c / FWHM$ and FWHM is the full width at half maximum of the Lorentzian.

Following \cite{Motta2017} and \cite{Casella2005}, we classified HBO-like QPOs as those found between a few mHz and \mbox{$\sim 60$ Hz} having a fractional rms between a few percent and $\sim 20 \%$, and falling on the rms-frequency track associated with HBO and HBO-like QPOs, as shown by \cite{Motta2017} (see Figure \ref{fig:lfqpos} and section \ref{subsec:coloura}). kHz QPOs were found between $\sim 400$ Hz and \mbox{$\sim 1500$ Hz} with fractional rms mostly lower than $\sim 20 \%$.

\subsection{Classification of triplets}
\label{sec:class}
The simultaneous occurrence of an HBO-like QPO, a lower kHz QPO and an upper kHz QPO in the PDS of an observation is referred to as a \textit{triplet}. We distinguish between three groups of triplets for this work: \textit{confirmed}, \textit{tentative} and \textit{incomplete} triplets.

Confirmed triplets are those containing QPOs that all have a significance at or above $3\sigma$ without any of the parameters used for the fitting of their Lorentzians having had to be constrained or fixed - as such, these triplets have the highest probability of being real, and not being the result of statistical fluctuations. Tentative triplets are those where one or more of the QPOs needed fitting parameters to be constrained in order for the QPO's significance to be adequately high - these triplets therefore have a lower probability of being real. We note that, for kHz QPOs, this was only done when a possible QPO could be distinguished clearly through visual inspection. Incomplete triplets only have $2$ out of the $3$ necessary QPOs present. Using \cite{IngramRPM2014}, these can still be used to place limits on the system using the RPM, and so they are included in our analysis.

We found in total 9 confirmed triplets, 15 tentative triplets and 7 incomplete triplets. Observation 90408-01-01-03 contains two \mbox{LF QPOs} that are both good candidates for an HBO, and so we include both these possible triplets for completeness. For the incomplete triplets, we only have cases where either the lower or upper kHz QPO is missing in observations. In all cases where the HBO-like QPO was missing, we included a Lorentzian with a starter width and frequency equal to the respective average characteristics calculated from the confirmed triplets. The width was then kept constrained, while the centre frequency was free to vary so as to see whether the significance of the feature would reach a sufficiently high value. In all these cases significant QPOs were found, and these former incomplete triplets are therefore rather grouped with the tentative triplets. The confirmed and tentative triplets found can be seen in Table \ref{table:tab1}, while the incomplete triplets are found in Table \ref{table:tab2}.

\section{Results}
\label{sec:results}
We produced the light curves, the HIDs and the frequency-rms diagrams for the LF QPOs and kHz QPOs in order to analyse \mbox{4U1608-52}, whereafter we used the RPM to calculate the spin $a$, mass $M$ and emission radius $r$ using the various groups of triplets found in our data.

\begin{figure*}
\centering
\includegraphics[width=1.0\textwidth]{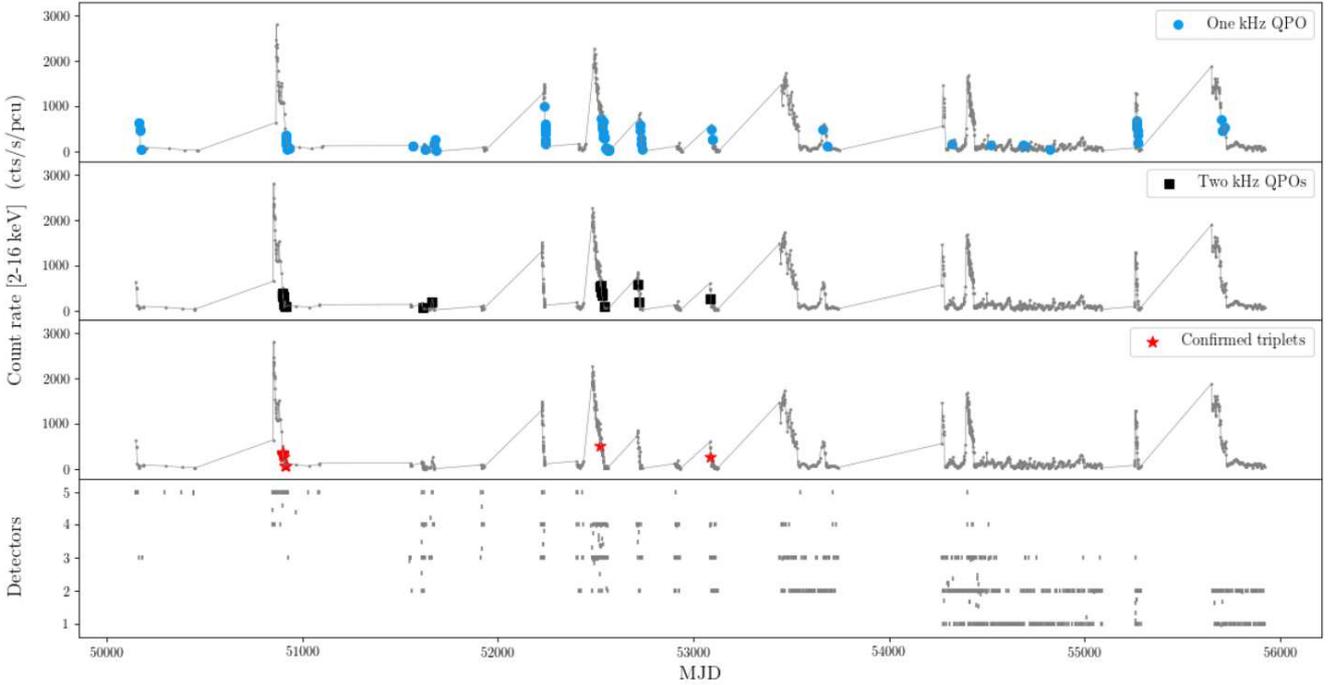}
\caption{\textbf{Top three panels:} The light curve of 4U1608-52, showing the source count rate of observations taken from March 1996 to December 2011, with each data point representing an observation, is displayed in each of the top three panels. The light blue circles in the top panel are observations containing one or more kHz QPOs, black squares in the second panel are observations containing two or more kHz QPOs (they comprise of both the confirmed and tentative triplets) and red stars in the third panel represent observations containing confirmed triplets. Red starred observations are a subset of the observations denoted by black squares, which in turn is a subset of the blue circular observations. A colour version is available online. \textbf{Bottom panel:} The number of PCU detectors working in each observation. When the number if PCUs is not an integer, it means that one or more PCUs turned on/off during the observation.} 
\label{fig:lc}
\end{figure*}

\subsection{Colour analysis and general behaviour}
\label{subsec:coloura}
The top three panels in Figure \ref{fig:lc} shows the light curve (the source count rate and date of each observation) of 4U1608-52, covering roughly 15 years from March 1996 to December 2011. Coloured data points give an indication of where kHz QPOs and confirmed triplets were found - light blue observations contain one or more kHz QPOs, black observations contain two kHz QPOs and red observations contain confirmed triplets. The black data points comprise of the tentative and confirmed triplets put together, so in all cases where a black dot is not overlapped by a red one, the observation contains a tentative triplet. The bottom panel in Figure \ref{fig:lc} shows the number of PCU detectors used for each observation. It is noted that as time progresses there is a tendency for fewer detectors to be used.

\begin{figure}
\centering
\includegraphics[width=0.5\textwidth]{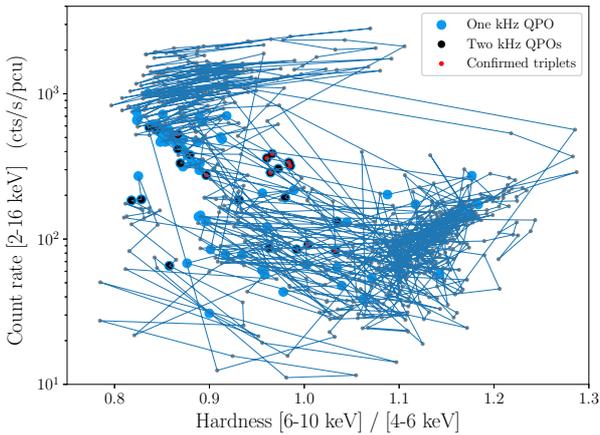}
\caption{The HID of 4U1608-52, with the source count rate on the y-axis and with each data point representing an observation. Similar to Figure \ref{fig:lc}, light blue observations contain one or more kHz QPOs, black observations contain two or more kHz QPOs (and comprise of both the confirmed and tentative triplets), while red observations only comprise of confirmed triplets. Consecutive observations are connected. A colour version is available online.} 
\label{fig:HID}
\end{figure}

The HID for our data is shown in Figure \ref{fig:HID}, showing the source count rate and hardness of each observation in our analysis. Consecutive observations are connected via thin lines, and the coloured data points have the same meaning as they did in Figure \ref{fig:lc}. The diagram shows a span of more than two orders of magnitude in source count rate, along with hysteresis patterns showing distinct excursions to very high luminosity, softer states, corresponding to the high luminosity phases present in the light curve in \mbox{Figure \ref{fig:lc}.} The source is shown to be harder during the rise of outbursts, and follows diagonal transitions (as noted by \citealt{Munoz2014}) to the brighter soft state. For 4U1608-52, the harder state corresponds to a fractional rms larger than $\sim 20$ \% while the soft state corresponds to a fractional rms smaller than $\sim 5$ \%, similar to what was found by \cite{Munoz2014}. Eventually, the luminosity (roughly tracked by the source count rate) decreases, preceding a reverse diagonal transition back to the harder state. Above the low-medium levels of luminosity, bright atolls often enter a soft, low-variability state where the correlation between variability and hardness breaks down (\citealt{Munoz2014}) and hysteresis stops. The HID of 4U1608-52 shows such bright excursions - they have a corresponding roughly constant fractional rms of \mbox{$\sim 2-3$ \%} (not immediately visible in the plots presented here).

\begin{figure}
\centering
\includegraphics[width=0.5\textwidth]{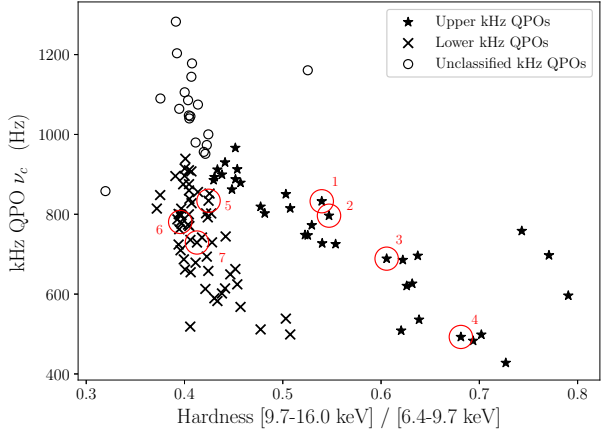}
\caption{The frequency and hardness of the kHz QPOs found in our dataset. Hardness $H_{\textrm{split}}$ is taken as the ratio of counts in two energy bands $C$ and $D$ as $H_{\textrm{split}} = C/D$, where $C$ is the band stretching between \mbox{$9.7-16.0$ keV} and $D$ stretches between $6.4-9.7$ keV for each observation. Stars denote upper kHz QPOs, crosses denote lower kHz QPOs and open circles denote either of the two in cases where it is hard to distinguish between them based solely on the figure. Red encircled points denote those kHz QPOs that form part of our set of incomplete triplets - the numbering aids the identification of the corresponding observations in Tables \ref{table:tab2} and \ref{tab:resultstab2}. A colour version is available online.}
\label{fig:hfqpoSplitOurs}
\end{figure}

In order to classify between lower and upper kHz QPOs in cases where only one of them is present in an observation, we plotted their centre frequencies $\nu_c$ against a hardness measure. \cite{Belloni2007} used similar plots and found that kHz QPOs generally group themselves into two tracks. QPOs along the top track of their plot were found to be upper kHz QPOs, while those along the bottom track were lower kHz QPOs - we use this scheme to classify our kHz QPOs. Hardness $H_{\textrm{split}}$ is here defined as $H_{\textrm{split}} = C/D$, where $C$ is the number of counts in the energy band stretching from \mbox{$9.7-16.0$ keV} and $D$ is the number of counts in the band \mbox{$6.4-9.7$ keV} in an observation - these ranges were taken from \cite{softhard}, as they proved to separate the different groups of QPOs best for our purposes. Our results are shown in Figure \ref{fig:hfqpoSplitOurs}. Stars denote upper kHz QPOs and crosses signify lower kHz QPOs. Open circles are those that do not with certainty fall on either of the branches, and classification is therefore withheld. The kHz QPOs encircled and numbered in red correspond to those that form part of the incomplete triplets - in this way, Figure \ref{fig:hfqpoSplitOurs} is used to classify the missing kHz QPOs in these triplets. The numbers are used to identify the corresponding observations (see Table \ref{table:tab2}). We note two exceptions to this classification - numbers 4 and 7 in Figure \ref{fig:hfqpoSplitOurs} - where we believe that usage of the figure leads to a misclassification of kHz QPOs. The reasons for this are discussed in \mbox{section \ref{subsec:rpmresults}.}


\begin{figure}
\centering
\includegraphics[width=0.5\textwidth]{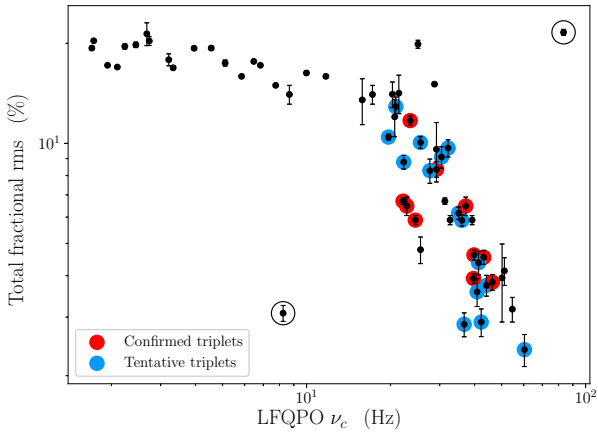}
\caption{The total fractional rms and frequency of all significant \mbox{LF QPOs} found in the dataset, with each data point representing one observation per LF QPO. The track stretching from the upper left to bottom right constitute HBO-like QPOs, the bottom outlier (encircled) is most probably an FBO-like QPO and the top outlier (also encircled) is a hecto hertz QPO. The red data points are those HBO-like QPOs that form part of our confirmed triplets, while the blue data points form part of the tentative triplets. A colour version is available online.} 
\label{fig:lfqpos}
\end{figure}

\begin{figure}
\centering
\includegraphics[width=0.5\textwidth]{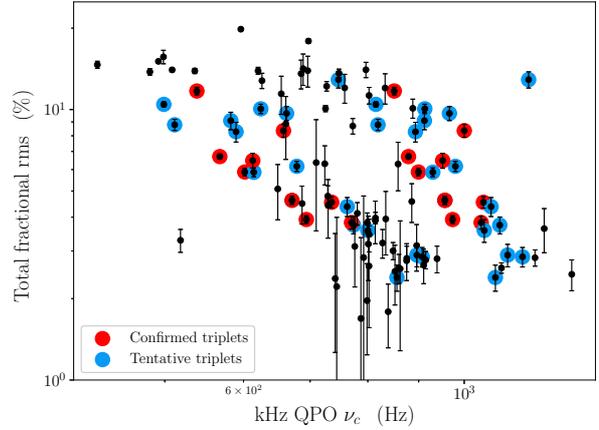}
\caption{The total fractional rms and frequency of all significant kHz QPOs found in the dataset, with each data point representing one observation per kHz QPO. The red data points are those QPOs that form part of the confirmed triplets, while the blue data points form part of the tentative triplets. A colour version is available online.} 
\label{fig:hfqpos}
\end{figure}

We next plot the rms-frequency diagrams for the LF QPOs and the kHz QPOs found in our data, as is shown in Figure \ref{fig:lfqpos} and \ref{fig:hfqpos}, respectively. When fitting these, constrained widths for QPOs were allowed if that resulted in them having a sufficiently high significance. Red and blue data points show those QPOs that form part of either our confirmed or tentative triplets, respectively. The frequency-rms diagram depicting the LF QPOs in our dataset (see Figure \ref{fig:lfqpos}) shows similar features to Figure 2 from the work of \cite{Motta2017}, with LF QPOs spanning similar frequency and rms ranges. From their LF QPO classification we can identify the QPOs forming part of the curve stretching from the upper left to the lower right as HBO-like QPOs. The top outlier (encircled) possibly represents a hectohertz QPO \citep{altamirano2008}. Following from \cite{Motta2017}'s scheme, the bottom outlier (also encircled) is most probably an FBO-like QPO. Furthermore, it can be seen from Figure \ref{fig:lfqpos} that QPO frequency increases as rms decreases along outbursts (also noted by \citealt{Munoz2014} and many others), corresponding to predictions by the Lense-Thirring effect whereby the HBO (or HBO-like) QPO's frequency depends on the outer radius of the precessing optically thin, geometrically thick inner accretion flow - the smaller the radius, the higher the frequency. Figure \ref{fig:hfqpos}, showing the frequency-rms diagram of the kHz QPOs found in our dataset, shows features similar to those in Figure 5 of \mbox{\cite{Motta2017}} - two tracks, predominantly falling in the rms range \mbox{$2 - 20$ \%.} Using Figure \ref{fig:hfqpoSplitOurs}, we find that the top kHz QPO track mostly corresponds to upper kHz QPOs, while the bottom track corresponds mostly to lower kHz QPOs. Note that the error bars on the observations along the lower part of the lower branch appear large mostly due to the logarithmic scaling of the axis. 

\subsection{Relativistic precession model}
\label{subsec:rpmresults}
\begin{figure*}
\centering
\includegraphics[width=0.7\textwidth]{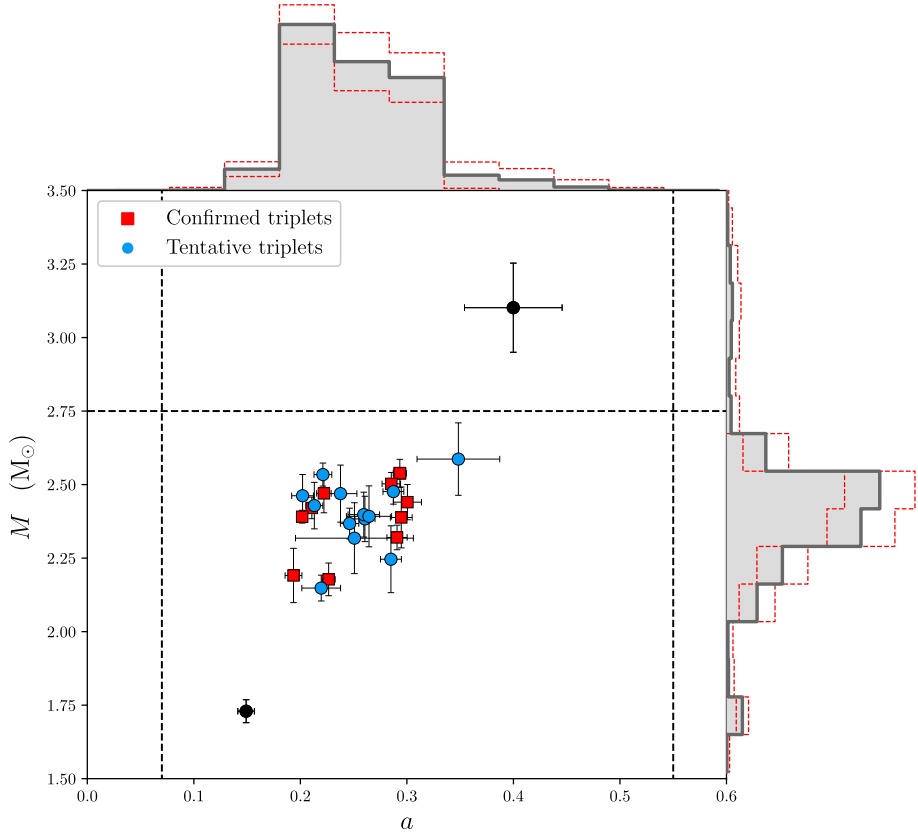}
\caption{The mass and spin estimates of both the confirmed (red squares) and tentative (blue circles) triplets, as predicted by the RPM. The black data points denote tentative triplets containing QPOs with Q-factors 4-5 times larger than that of other triplets - these can therefore very possibly not be real. The marginal distributions of the spin and mass can be seen at the top and to the right of the figure, respectively, with the $1\sigma$ error on these quantities also given. The broken black lines indicate theoretical physical limits for the spin and the mass, taking into account current models of the equation of state and approximating the NS as both a hollow and a solid sphere. A colour version is available online.}
\label{fig:mass_spin_all1}
\end{figure*}

\begin{figure}
\centering
\includegraphics[width=0.50\textwidth]{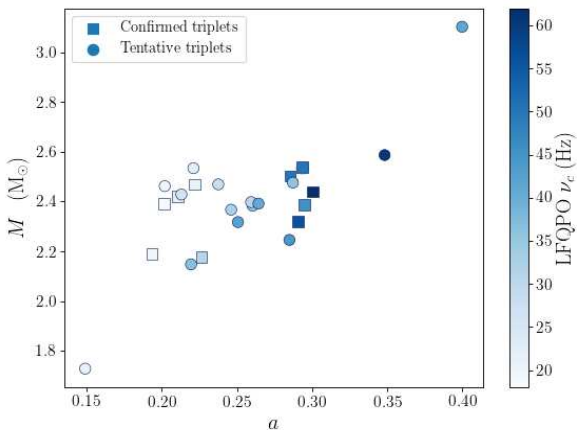}
\caption{The mass and spin estimates of the confirmed and tentative triplets as predicted by the RPM, coloured according to the LF QPO frequency. A colour version is available online.} 
\label{fig:mass_spin_lfqpo_all1}
\end{figure}

\begin{figure}
\centering
\includegraphics[width=0.50\textwidth]{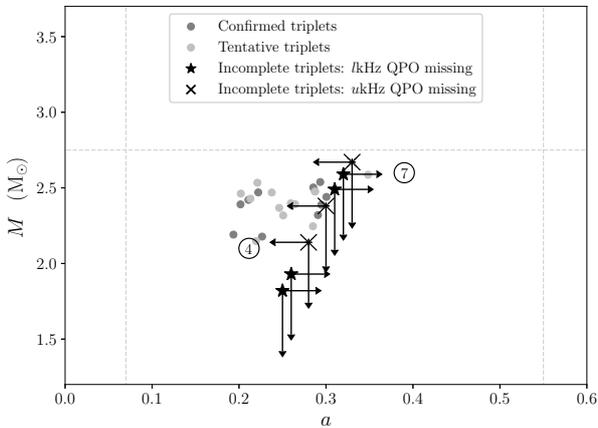}
\caption{The mass and spin estimates of the confirmed and tentative triplets along with the mass and spin constraints given by incomplete triplets either lacking an upper kHz QPO ($u$kHz QPO) or a lower kHz QPO ($l$kHz QPO).}
\label{fig:incomp}
\end{figure}

We used the RPM (see section \ref{sec:rpm}) to calculate the spin $a$, mass $M$ and emission radius $r$ of 4U1608-52 using each of our confirmed, tentative and incomplete triplets. The results are shown in Table \ref{tab:resultstab1} and \ref{tab:resultstab2}, and in \mbox{Figure \ref{fig:mass_spin_all1}}, \ref{fig:mass_spin_lfqpo_all1} and \ref{fig:incomp}.

\begin{table*}
\renewcommand{\arraystretch}{1.5}
\caption[Results11]{The spin, mass and emission radius of the confirmed and tentative triplets as determined with the RPM. The frequencies of the LF QPOs are also included so that information shown in Figure \ref{fig:mass_spin_lfqpo_all1} can be seen clearly.}
    \begin{tabular}{lcccc}
    \toprule
    \multicolumn{5}{c}{\textbf{Confirmed Triplets}} \\ \toprule
    \textbf{Obs ID}             & $\bm{a}$ & $\bm{M/\textrm{M}_\odot}$ & $\bm{r}$ & \textbf{LF QPO (Hz)} \\ \toprule      
    30062-02-01-00              & $0.293 \pm 0.006$         & $2.54 \pm 0.05$       & $5.54 \pm 0.04$     & $40^{+1}_{-1}$       \\
    30062-02-01-01-1            & $0.211 \pm 0.004$         & $2.42 \pm 0.04$       & $5.99 \pm 0.03$     & $24.5^{+0.5}_{-0.6}$       \\
    30062-02-01-01-3            & $0.202 \pm 0.003$         & $2.39 \pm 0.02$       & $6.14 \pm 0.02$     & $22.2^{+0.4}_{-0.4}$       \\
    30062-02-01-02-1            & $0.30 \pm 0.01$       & $2.44 \pm 0.06$       & $5.37 \pm 0.06$     & $46^{+2}_{-3}$       \\
    30062-01-01-00-1            & $0.291 \pm 0.009$         & $2.32 \pm 0.04$       & $5.54 \pm 0.04$     & $43^{+2}_{-2}$       \\
    30062-01-02-01              & $0.222 \pm 0.007$         & $2.47 \pm 0.07$       & $6.13 \pm 0.09$     & $23.5^{+0.5}_{-0.6}$       \\
    30062-01-02-05-1            & $0.227 \pm 0.006$         & $2.18 \pm 0.06$       & $5.99 \pm 0.05$     & $29.2^{+0.9}_{-0.9}$       \\
    70059-01-20-00              & $0.285 \pm 0.009$         & $2.50 \pm 0.04$       & $5.53 \pm 0.04$     & $40^{+1}_{-1}$       \\
    90408-01-01-03              & $0.194 \pm 0.008$         & $2.19 \pm 0.09$       & $6.18 \pm 0.08$     & $23^{+1}_{-1}$       \\
    ~                           & $0.30 \pm 0.01$       & $2.4 \pm 0.1$       & $5.80 \pm 0.09$     & $37^{+2}_{-2}$       \\ \toprule
    \multicolumn{5}{c}{\textbf{Tentative triplets}} \\ \toprule
    30062-02-01-01-4            & $0.29 \pm 0.01$       & $2.48 \pm 0.04$       & $5.75 \pm 0.05$     & $36^{+2}_{-2}$       \\
    30062-01-01-00-2            & $0.29 \pm 0.01$       & $2.2 \pm 0.1$       & $5.53 \pm 0.09$     & $44^{+2}_{-2}$       \\
    30062-01-01-02              & $0.261 \pm 0.009$         & $2.38 \pm 0.08$       & $5.70 \pm 0.06$     & $35^{+2}_{-2}$       \\
    30062-01-01-03              & $0.20 \pm 0.01$       & $2.46 \pm 0.07$       & $6.3 \pm 0.1$     & $20^{+1}_{-1}$       \\
    30062-01-02-05-2            & $0.213 \pm 0.008$         & $2.43 \pm 0.08$      & $5.9 \pm 0.1$     & $25.6^{+0.7}_{-0.7}$       \\
    50052-01-20-00              & $0.24 \pm 0.02$       & $2.5 \pm 0.1$       & $5.93 \pm 0.09$     & $28^{+2}_{-2}$       \\
    50052-01-20-01              & $0.26 \pm 0.02$       & $2.40 \pm 0.08$       & $5.96 \pm 0.08$     & $30^{+2}_{-2}$       \\
    50052-01-04-00              & $0.149 \pm 0.008$         & $1.73 \pm 0.04$       & $6.34 \pm 0.08$     & $21^{+1}_{-1}$       \\
    70058-01-37-00              & $0.35 \pm 0.04$       & $2.6 \pm 0.1$       & $5.0 \pm 0.2$     & $60^{+8}_{-9}$       \\
    70059-03-01-01              & $0.26 \pm 0.02$       & $2.4 \pm 0.1$       & $5.4 \pm 0.1$     & $41^{+4}_{-4}$       \\
    70059-03-02-00              & $0.25 \pm 0.06$       & $2.3 \pm 0.1$       & $5.4 \pm 0.2$     & $42^{+6}_{-16}$       \\
    70059-03-02-01              & $0.40 \pm 0.05$       & $3.1 \pm 0.2$       & $5.6 \pm 0.2$     & $41^{+4}_{-8}$       \\
    70059-01-26-00              & $0.246 \pm 0.009$         & $2.37 \pm 0.05$       & $5.78 \pm 0.05$     & $32^{+1}_{-1}$       \\
    80406-01-02-02              & $0.22 \pm 0.02$       & $2.15 \pm 0.04$       & $5.51 \pm 0.07$     & $37^{+4}_{-3}$       \\
    80406-01-03-03              & $0.221 \pm 0.008$         & $2.53 \pm 0.04$       & $6.18 \pm 0.05$     & $22^{+1}_{-1}$        \\ 
    
    \end{tabular}
    \label{tab:resultstab1}
\end{table*}

Figure \ref{fig:mass_spin_all1} shows the mass $M$ and spin $a$ predicted by each of our confirmed (red squares) and tentative (blue circles) triplets, with errors calculated as specified in section \ref{sec:rpm}. The two outlier triplets (in black: 50052-01-04-00 to the bottom and \mbox{70059-03-02-01} to the top) form part of the tentative triplets - we mark them as having the lowest probability of being real, as both contain upper kHz QPOs with Q-factors significantly higher than the other triplets here used (see Table \ref{tab:resultstab1}). We exclude these two outliers in further figures. The dashed black lines indicate theoretical upper and lower physical limits for the mass and spin of our system by making use of the equation of state (EoS) and approximating the NS as both a hollow and solid sphere, respectively.

To calculate the physical limits of the dimensionless spin parameter, we took the spin frequency of the NS to be $\nu = 619$ Hz (\citealt{gallo}) and the mass as $2.38 \hspace{1mm} \textrm{M}_\odot$, the average of the masses predicted by using the confirmed triplets in the RPM (we used the first of the instances of 90408-01-01-03 listed in \mbox{Table \ref{tab:resultstab1},} as its LF QPO is slightly more significant than the alternative). Assuming that the radius $R$ of a NS falls between $7$ km and $15$ km \citep{WhyDisInNature} and approximating the NS as a hollow and solid sphere, respectively, we calculated the maximum spin, for a hollow sphere with \mbox{$R=15$ km}, as $a_{\textrm{max}} = 0.55$ using the definition of the dimensionless spin parameter given in section \ref{sec:rpm}. The minimum spin, for a solid sphere with \mbox{$R=7$ km}, was found to be $a_{\textrm{min}} = 0.07$. These limits are indicated in Figure \ref{fig:mass_spin_all1}. Upper and lower limits for the allowable mass of a NS were obtained by making use of the predictions of the EoS (\citealt{WhyDisInNature}) - the upper limit being $\sim 2.75 \hspace{1mm} \textrm{M}_\odot$ and the lower limit being $\sim 0.35 \hspace{1mm} \textrm{M}_\odot$ (the upper limit is indicated in Figure \ref{fig:mass_spin_all1}). These limits represent the most extreme allowed physical limits, and give an idea of where acceptable mass and spin values should lie. It can be seen that, in all but one case, the RPM results in mass and spin values that fall within these limits. Furthermore, except from the two outlier cases (in black), all the values cluster roughly between spin values of $0.19 < a < 0.35$ and mass values of $2.15 < M/\textrm{M}_{\odot} < 2.6$. Doing the reverse and assuming a solid sphere, a spin frequency of $\nu = 619$ Hz, and using each of the confirmed triplets' spin-mass parameter value pairs in turn (see Table \ref{tab:resultstab1}) we calculate the radius $R$ of the neutron star to fall in the range $11.0 < R < 14.6$ km.

The marginal distributions of the spin and mass are supplied in the form of histograms at the top and to the right of Figure \ref{fig:mass_spin_all1}. We calculated these distributions by using a Monte Carlo approach. For each triplet, values for the frequencies of the LF QPO, lower kHz QPO and upper kHz QPO were drawn from Gaussian distributions having the respective fitted frequency values (see Table \ref{table:tab1}) as their means, and the average error on each of these quantities as their standard deviations. The numerous combinations of these three frequency values were then used to calculate corresponding values for $a$, $M$ and $r$ with the RPM equations. This was done for each of our confirmed and tentative triplets, after which all results for $a$ and $M$ were binned separately in order to obtain the marginal distributions. The standard deviation in each bin was used as the $1\sigma$ error, indicated in red dashed lines on the histograms in the figure.

Next, we plotted the calculated values of $a$ and $M$ of the confirmed and tentative triplets, colour coded according to their \mbox{LF QPO} frequencies, in Figure \ref{fig:mass_spin_lfqpo_all1}. Lighter colours indicate lower frequencies while darker colours indicate higher frequencies, and squares and circles denote confirmed and tentative triplets, respectively. It can be seen that lighter colours tend towards the bottom left of the figure, and that moving upwards and right brings about darker colours.

Lastly, Figure \ref{fig:incomp} shows, in different shades of grey, the spin and mass of the confirmed and tentative triplets along with the theoretical physical limits these parameters are allowed to have, as was also shown in Figure \ref{fig:mass_spin_lfqpo_all1}. Added to this, however, are the limits on $a$ and $M$ predicted by each of our incomplete triplets (see Table \ref{tab:resultstab2}), with starred triplets being those lacking lower kHz QPOs and crossed triplets those that lack upper kHz QPOs. When the kHz QPOs of triplets 4 and 7 are classified using Figure \ref{fig:hfqpoSplitOurs}, we find triplet 4's limits to be $M \leq 4.66 \textrm{M}_\odot$ and $a \geq 0.45$, while triplet 7 results in limits of $M \leq 1.16$ and $a \leq 0.18$. These limits make triplets 4 and 7 extreme outliers, and is a likely indication that classifying kHz QPOs using \mbox{Figure \ref{fig:hfqpoSplitOurs}} might result in the misclassification of some of the isolated peaks.
The low value of triplet 4's detected kHz QPO of \mbox{$493$ Hz} places further doubt on its classification as an upper kHz QPO by Figure \ref{fig:hfqpoSplitOurs}. However, classifying triplet 4 and 7's kHz QPOs as a lower kHz QPO and upper kHz QPO (see Table \ref{table:tab2}), respectively, results in much more feasible limits on their $a$ and $M$ values, as shown in Figure \ref{fig:incomp} and Table \ref{tab:resultstab2}. We note that kHz QPOs have been classified also based on their phase lags (see e.g. \citealt{nowak1999} on the calculation of lags), which in some sources show a dependence on the type of kHz QPO detected (i.e. upper or lower)  - see e.g. \cite{avellar2013, avellar2016}, \cite{barret2013} and \cite{peille2015} for their work on NS QPO lags, and \cite{mendex2013} for their work on BH QPO lags in this regard. However, since the classification based on the phase lags suffers from uncertainties, a careful and detailed study of the lags associated to the QPOs in 4U1608-52 would be required to obtain reliable  and solid results. Such a study is however beyond the scope of this paper, and is left to a future work.

\begin{table*}
\renewcommand{\arraystretch}{1.5}
\caption[Results12]{Limits placed on the spin, mass and emission radius by making use of the incomplete triplets, as calculated with the RPM. Numbers assigned to observations in the second column correspond to the numbers similarly assigned in Table \ref{table:tab2} and Figure \ref{fig:hfqpoSplitOurs}.}
    \begin{tabular}{lccccc}
    \toprule
    \multicolumn{6}{c}{\textbf{Incomplete Triplets}} \\ \toprule
    \textbf{Obs ID}     &  \#     & $\bm{a}$ & $\bm{M/\textrm{M}_\odot}$ & $\bm{r}$ & \textbf{LF QPO (Hz)} \\ \toprule
    70069-01-03-01  & 1 & $\geq 0.26$  &  $\leq 1.93$  &  $\geq 7.34$  &  $20.7^{+0.8}_{-0.7}$  \\
    70069-01-03-06  & 2 & $\geq 0.25$  &  $\leq 1.82$  &  $\geq 7.86$  &  $17.2^{+0.4}_{-0.3}$  \\
    50052-01-24-00  & 3 & $\geq 0.31$  &  $\leq 2.49$  &  $\geq 7.02$  &  $21.4^{+0.4}_{-0.4}$  \\
    60052-03-01-00  & 4 & $\leq 0.28$  &  $\leq 2.14$  &  $\geq 5.71$  &  $28.7^{+0.8}_{-1.2}$  \\
    95334-01-03-11  & 5 & $\leq 0.30$  &  $\leq 2.38$  &  $\geq 5.29$  &  $50^{+3}_{-2}$  \\
    70059-03-02-03  & 6 & $\leq 0.33$  &  $\leq 2.67$  &  $\geq 5.16$  &  $51^{+4}_{-3}$  \\
    95334-01-03-05  & 7 & $\geq 0.32$  &  $\leq 2.59$  &  $\geq 7.2$   &  $25.6^{+0.7}_{-0.6}$  \\   
    
    \end{tabular}
    \label{tab:resultstab2}
\end{table*}

\section{Discussion}
\label{sec:disc}
We have examined all publicly available RXTE PCA data on the NS LMXB 4U1608-52 taken between March 1996 and December 2011 with the purpose of finding observations containing QPOs - particularly those containing QPO triplets to be used in the testing of the RPM. In total we found 9 confirmed triplets, 15 tentative triplets and 7 incomplete triplets (see Table \ref{table:tab1} and \ref{table:tab2}).

High luminosity phases along 4U1608-52's light curve (top three panels, \mbox{Figure \ref{fig:lc})} can be seen clearly, and it is noted that observations are more frequent on the downward slopes than on the rise. This is due to a combined effect: the rise time of the high luminosity phase being shorter than the fall, as well as activity only being spotted when the bursts make the sources adequately bright - it will therefore only be monitored predominantly from such a point onwards. It is further noted that as time progresses, the number of detectors typically used per observation (bottom panel, Figure \ref{fig:lc}) steadily decreases. We ascribe the lack in detections of any triplets after an MJD of roughly 53200 to the steady decrease in total source count rate per observation due to the decrease in the number of PCUs used.


Our results, when applying the RPM to our confirmed, tentative and incomplete triplets, are summarised in Table \ref{tab:resultstab1} and \ref{tab:resultstab2} and Figure \ref{fig:mass_spin_all1}, \ref{fig:mass_spin_lfqpo_all1} and \ref{fig:incomp}. We note that our estimates are not completely consistent, as can be seen in Figure \ref{fig:mass_spin_all1}, and we identify several reasons at the base of these inconsistencies in section \ref{sec:cav}. It is, however, very reassuring that the vast majority of estimates fall within physically allowable limits (indicated in broken lines in Figure \ref{fig:mass_spin_all1}), even considering that these limits are slightly unrealistic (the NS being approximated as both a hollow and solid sphere, respectively). Additionally, apart from the two outliers (in black, discussed in section \ref{subsec:rpmresults}), most estimates cluster together - a further reassuring sign. Our mass estimates fall towards the higher end of the allowable NS mass ranges, and are compatible with many of the EoS models - however, if our measurements are correct, we would be ruling out some of the more exotic models investigated in \cite{WhyDisInNature}. We further note that in the case of our mass measurements being correct (i.e. falling in the range $2.15 < M/\textrm{M}_\odot < 2.6$), the NS in 4U1608-52 would be one of the heaviest known to date, with the current highest-mass NS being PSR J2215+5135 \citep{Linares2018} at $2.27^{+0.17}_{-0.15} \hspace{1mm} \textrm{M}_\odot$, closely followed by PSR J0348-0432 with a mass of $2.01 \pm 0.04 \hspace{1mm} \textrm{M}_\odot$ \citep{antoniadis2013}. 

Based on the current knowledge of compact objects, heavy NSs can only be produced by stiff EoSs, and it is therefore believed that the softest EoSs are likely unrealistic \citep{horvath2017}. 
However, the fact that stiffer EoSs are needed to explain such heavy NSs poses the problem for alternative models of the internal composition, like the hyperon model \citep{horvath2017}. Hyperons, existing as part of the natural baryonic sector, have been included in some dense matter studies of NSs for the last 40 years. Their inclusion softens stiff EoSs considerably, precluding precisely the higher masses of NSs - this is called the ``hyperon puzzle" \citep{horvath2017, ozel2016}. Including hyperons, EoSs barely accomodate NSs reaching $2 \textrm{M}_{\odot}$. 
For a NS as large as that in 4U1608-52, therefore, other alternatives would have to be considered. One solution would be the possible suppression of hyperons due to some yet unknown mechanism; an alternative explanation would be that that there is enough repulsion from them to explain these effects \citep{ozel2016, horvath2017}.

\cite{rpm2} already noticed that the RPM tends to  yield high masses. As an explanation they proposed that the HBO associated to the kHz QPOs in triplets is the second harmonic of the real Lense-Thirring frequency, therefore the Lense-Thirring frequency would correspond to half of the centroid frequency measured in the PDS (see alsot \citealt{Motta2017}). Taking this into account and assuming that the HBO frequency we measured corresponds to the second harmonic of a non-detected fundamental frequency, we find that the spin and mass values for 4U1608-52 would fall in the ranges $0.10 < a < 0.19$ and $1.97 < M/\textrm{M}_\odot < 2.33$. This is shown in Figure \ref{fig:appB}. It is noted that even with this lowered mass range, the mass is still large enough to cause problems for stiff EoSs model in light of the ``hyperon puzzle". If the mass were to lie towards the lower end of this range, some of these more exotic models of the EoS would still (even though only barely) apply \citep{WhyDisInNature, horvath2017}.


Figure \ref{fig:mass_spin_lfqpo_all1} shows estimates corresponding to lower LF QPO frequencies tending to lower spin and mass values, while those associated with higher frequencies tend towards higher spins and masses. Seeing as all QPO frequencies reach their highest allowed values at the ISCO, we know that those estimates with higher frequencies were found closer to the ISCO than those with lower frequencies. The RPM is an increasingly better approximation of the rigid precession model (\citealt{Ingram2009}) the closer to the ISCO the LF QPO originates from. As the rigid precession model features a better modelling of what is believed to be the true motion of the accretion disk than the RPM, higher frequency $(a,M)$ parameter pairs are therefore more likely to be a better representation of the real system (see section \ref{sec:cav} for a further elaboration). Seeing as the estimate relating to the highest-frequency LF QPO results in the largest spin and mass values, all other lower-frequency estimates serve as lower limits to these values. Based on this reasoning, if one further assumes that this same LF QPO is emitted at the ISCO, the ISCO radius can be calculated and used as an upper limit to the radius of the NS. Using the relation between the $u$kHz QPO's centroid frequency and the radius of the ISCO ($r_{ms}$) given in \cite{bookvanderklis}, we find $r_{ms} = 18.6$ km.

In Figure \ref{fig:incomp}, it is once again reassuring to see that the estimates (and their imposed limits) given by the incomplete triplets seem to cluster with those predicted by the confirmed and tentative triplets.

\subsection{Caveats}
\label{sec:cav}
We outline three caveats to this study and to the results obtained that should be kept in mind when interpreting findings involving the RPM.

Firstly, assuming the correctness of the RPM, the most obvious source of biases in our spin and mass measurements resides in the QPO classification - given the intrinsic difficulty in classifying NS QPOs (NS PDSs are more complex than those of BHs, and QPOs are less clear), it is possible that some of our triplets involve QPOs that are not HBO-like QPOs. This implies that they could return untrustworthy results. Furthermore, as mentioned in \mbox{section \ref{sec:disc}}, it is possible that incorrectly classified kHz QPOs in incomplete triplets could lead to further erroneous results. 

Secondly, the RPM constitutes a simplified modeling of what is a certainly more complex behaviour of an ensemble of particles, like the accretion flow is. A more accurate description of the origins of QPOs around compact objects is given by the rigid precession model \cite{Ingram2009}, which is still a work in progress and is itself currently limited to LF QPOs. The RPM, however, is a good approximation of the rigid precession model (see section \ref{sec:intro}) when used with QPOs produced close to the ISCO (see \citealt{rigidpmMotta}). Therefore, assuming that our triplets all consist of correctly classified QPOs, it might be possible that our results are flawed due to the use of QPOs produced far from the ISCO. Indeed, we show that proximity to the ISCO has a visible effect on our estimates (see \mbox{Figure \ref{fig:mass_spin_lfqpo_all1}).}


Lastly, it is of course possible that neither the RPM nor the rigid precession model constitute a good description of the real properties of the accretion flow. Further investigations will clarify this, however we are inclined to believe that this will not be the case, given that these models successfully describe observational facts in a quantitative way and in a number of different contexts \citep{stella1999, done1, IngramRPM2014, m2014, Motta2014b}.

\section{Summary and Conclusions}
\label{sec:summ}
We have analysed all the available RXTE/PCA data on the NS 4U1608-52 and produced the light curve, the HID and the frequency-rms diagrams for the LF QPOs and kHz QPOs found in the data. The HID shows the hysteresis patterns that we typically expect for atoll sources, along with bright soft-state excursions that reach bright atoll luminosities. The frequency-rms diagrams show the characteristic tracks associated with HBO-like QPOs and upper and lower kHz QPOs in the literature. We found in total 9 confirmed QPO triplets, 15 tentative triplets and 7 incomplete triplets with which to test the RPM.

By solving the RPM using the triplets we found, we obtained a set of spin-mass estimates, which mostly cluster in ranges of $0.19 < a < 0.35$ and $2.15 < M/\textrm{M}_\odot < 2.6$, well within the physically allowed spin and mass limits determined based on the NS EoSs. 
While our spin and mass estimates are not fully consistent, we identified and discussed reasons for the presence of both scatter in our estimates and the presence of outliers. In particular, we showed that these values tend to increase with proximity to the ISCO as the RPM becomes an increasingly good approximation of the real behaviour of matter in proximity to the NS for decreasing radii. 
If our estimates are correct, 4U1608-52 would be one of the heaviest NSs known to date. Our results would rule out some of the more exotic models of the EoS. 

We plan to extend this study to include a larger set of sources in order to better test the RPM on NS systems. 



\section*{Acknowledgments}
\label{sec:acknowledgments}
LdB acknowledges support from the Rhodes Trust and Christ Church, and would especially like to thank Kaldi posthumously for his alleged discovery of the effects of coffee berries in 9th-century Ethiopia. SEM \mbox{acknowledges} the Science and Technology Facilities Council (STFC) for financial support. We thank the referee for their thoughtful and helpful comments in reviewing this work.

\bibliographystyle{mnras}
\bibliography{main}



%

\appendix
\onecolumn

\begin{landscape} \section{Confirmed, tentative and incomplete triplets} \label{AppA}
	
    \vspace{-0.2cm}
    
	\captionof{table}[Triplets]{\textbf{Confirmed Triplets}: All observations in which all three constituent QPOs of the RPM (HBO, lower kHz QPO and upper kHz QPO] could be found. This is our most certain and prime source of triplets for 4U1608-52; none of the component parameters used in the spectral fitting of these observations were fixed or constrained in order to improve the significance or the Q factor of a potential QPO. \textbf{Tentative Triplets}: Observations which, with the help of some parameters being constrained during spectral fitting, also contained all three QPOs. A red starred value for a Q factor means that the width of the QPO with said Q factor was fixed in order for it to either have a qualifying Q factor or a qualifying significance, or both.}
	\small%
	\noindent

	\begin{center}
    \renewcommand{\arraystretch}{1.5}
    \begin{tabularx}{1.24\textwidth}{lccccccccccc}
    \toprule
    
    \multicolumn{12}{c}{\textbf{Confirmed triplets}} \\ \toprule
    
    \textbf{Obs ID} & \textbf{MJD} & \textbf{$l$kHz QPO (Hz)} & \textbf{Q factor} & \textbf{Sign.} & \textbf{$u$kHz QPO (Hz)} & \textbf{Q factor} & \textbf{Sign.} & \textbf{LF QPO (Hz)} & \textbf{Amplitude}   & \textbf{Q factor} & \textbf{Sign.} \\ \toprule
    
       30062-02-01-00           &    50896.91833 &    $671^{+2}_{-2}$    &    $28 \pm 3$  &    11.5      &    $956^{+10}_{-10}$    &    $8 \pm 1$   &    8.8       &    $40^{+1}_{-1}$    &  $\num{1.1e-03}^{+\num{1e-04}}_{-\num{1e-04}}$   &    $2.6 \pm 0.5$   &    9.9       \\
       
       30062-02-01-01-1         &    50897.58019 &    $601.7^{+0.6}_{-0.6}$    &    $49 \pm 5$  &    20.5      &    $899^{+7}_{-7}$    &    $7 \pm 2$   &    7.5       &    $24.5^{+0.5}_{-0.6}$    & $\num{3e-04}^{+\num{1e-04}}_{-\num{1e-04}}$    &    $9 \pm 5$   &    3.3       \\
       
       30062-02-01-01-3         &    50897.58019 &    $567.9^{+0.8}_{-0.9}$    &    $27 \pm 4$  &    13.1      &    $879^{+4}_{-4}$    &    $8 \pm 1$   &    11.6      &    $22.2^{+0.4}_{-0.4}$    &    $\num{3e-04}^{+\num{2e-04}}_{-\num{1e-04}}$ &    $8 \pm 4$   &    3.1       \\
       
       30062-02-01-02-1         &    50898.58019 &    $770.4^{+0.8}_{-0.8}$    &    $37 \pm 3$  &    17.8      &    $1040^{+13}_{-12}$   &    $9 \pm 3$   &    4.7       &    $46^{+2}_{-3}$    &    $\num{1.1e-03}^{+\num{3e-04}}_{-\num{2e-04}}$ &    $2 \pm 1$   &    4.9       \\
       
       30062-01-01-00-1         &    50899.51955 &    $735.8^{+0.4}_{-0.4}$    &    $50 \pm 3$  &    35.9      &    $1046^{+9}_{-10}$   &    $10 \pm 4$   &    4.7       &    $43^{+2}_{-2}$    &    $\num{5e-04}^{+\num{2e-04}}_{-\num{2e-04}}$ &    $4 \pm 2$   & 3.0          \\
       
       30062-01-02-01           &    50906.64556 &    $538^{+13}_{-20}$    &    $5 \pm 3$   &    3.1       &    $850^{+7}_{-7}$    &    $10 \pm 3$   &    5.4       &    $23.5^{+0.5}_{-0.6}$    &    $\num{2.0e-03}^{+\num{6e-04}}_{-\num{5e-04}}$ &    $4 \pm 2$   &    4.1       \\
       
       30062-01-02-05-1         &    50914.17628 &    $658^{+2}_{-2}$    &    $25 \pm 4$  &    10.2      &    $1000^{+13}_{-14}$   &    $10 \pm 5$  &    3.7       &    $29.2^{+0.9}_{-0.9}$    &    $\num{4.1e-03}^{+\num{5e-04}}_{-\num{5e-04}}$ &    $2.1 \pm 0.4$   &    8.5       \\
       
       70059-01-20-00           &    52524.10185 &    $694^{+2}_{-3}$    &    $17 \pm 2$  &    9.9       &    $974^{+7}_{-7}$    &    $10 \pm 2$  &    7.8       &    $40^{+1}_{-1}$    &    $\num{8e-04}^{+\num{2e-04}}_{-\num{1e-04}}$ &    $2.0 \pm 0.5$   &    6.1       \\
       
       90408-01-01-03           &    53088.39759 &    $613^{+1}_{-1}$    &    $45 \pm 8$  &    9.8       &    $951^{+22}_{-21}$    &    $7 \pm 2$   &    3.9       &    $23^{+1}_{-1}$    &    $\num{7e-04}^{+\num{4e-04}}_{-\num{2e-04}}$ &    $6 \pm 4$   &    4.3       \\
       
         &   53088.39759  &    $613^{+1}_{-1}$    &    $45 \pm 8$  &    9.8       &    $951^{+22}_{-21}$    &      $7 \pm 2$     &    3.9      &   $37^{+2}_{-2}$     &   $\num{1.7e-03}^{+\num{6e-04}}_{-\num{5e-04}}$    &  $4 \pm 2$   &  3.5 \\ \toprule
       
    \multicolumn{12}{c}{\textbf{Tentative triplets}} \\ \toprule
    
       30062-02-01-01-4         &    50897.58019 &    $614^{+1}_{-1}$    &    $25 \pm 2$  &    18.2      &    $930^{+7}_{-7}$    &    $10 \pm 2$  & 7.0          &    $36^{+2}_{-2}$    &    $\num{2.2e-03}^{+\num{2e-04}}_{-\num{2e-04}}$ &    \textcolor{red}{2.1*}   &    11.0      \\
       
       30062-01-01-00-2         &    50899.51955 &    $773.9^{+0.6}_{-0.6}$    &    $54 \pm 3$  &    29.0      &    $1086^{+32}_{-29}$   &     \textcolor{red}{5.0*}   &    4.6       &    $44^{+2}_{-2}$    &    $\num{7e-04}^{+\num{2e-04}}_{-\num{2e-04}}$ &    $4 \pm 1$   &    4.4       \\
       
       30062-01-01-02           &    50901.34889 &    $678.8^{+0.3}_{-0.4}$    &    $48 \pm 5$  &    24.2      &    $980^{+17}_{-17}$    &    $8 \pm 2$   &    4.7       &    $35^{+2}_{-2}$    &    $\num{2.2e-03}^{+\num{2e-04}}_{-\num{2e-04}}$ &     \textcolor{red}{2.1*}   &    9.7       \\
       
       30062-01-01-03           &    50902.52537 &   $499^{+18}_{-15}$    &    $6 \pm 3$   & 3.0          &    $815^{+6}_{-6}$    &    $6.3 \pm 0.8$   &    11.4      &    $20^{+1}_{-1}$    &    $\num{1.8e-03}^{+\num{3e-04}}_{-\num{3e-04}}$ &     \textcolor{red}{2.1*}   &    6.3       \\
       
       30062-01-02-05-2         &    50914.17628 &    $625^{+26}_{-26}$    &     \textcolor{red}{5.2*}   &    3.3       &    $913^{+9}_{-10}$    &    $10 \pm 3$  &    4.9       &    $25.6^{+0.7}_{-0.7}$    &    $\num{4.6e-03}^{+\num{4e-04}}_{-\num{4e-04}}$ &     \textcolor{red}{2.0*}   &    12.7      \\
       
       50052-01-20-00           &    51660.04616 &    $590^{+3}_{-4}$    &    $33 \pm 16$  & 4.0          &    $893^{+18}_{-18}$    &    $9 \pm 5$   &    3.1       &    $28^{+2}_{-2}$    &    $\num{3.9e-03}^{+\num{7e-04}}_{-\num{7e-04}}$ &     \textcolor{red}{2.0*}   &    5.3       \\
       
       50052-01-20-01           &    51660.29390 &    $583^{+6}_{-4}$    &    $21 \pm 11$  &    3.6       &    $912^{+13}_{-15}$    &     \textcolor{red}{13.6*}  &    3.9       &    $31^{+2}_{-2}$    &    $\num{4.7e-03}^{+\num{8e-04}}_{-\num{8e-04}}$ &     \textcolor{red}{2.0*}   &    5.8       \\
       
       50052-01-04-00           &    51614.05684 &    $747^{+19}_{-19}$    &     \textcolor{red}{4.8*}   &    5.9       &    $1161^{+9}_{-6}$   &     \textcolor{red}{42.6*}  &    3.4       &    $21^{+1}_{-1}$    &    $\num{4.8e-03}^{+\num{8e-04}}_{-\num{8e-04}}$ &     \textcolor{red}{2.6*}   &    6.2       \\
       
       70058-01-37-00           &    52525.14340 &    $855.8^{+0.6}_{-0.4}$    &    $156 \pm 52$ &    22.9      &    $1075^{+17}_{-20}$   &     \textcolor{red}{13.3*}  &    3.4       &    $60^{+8}_{-9}$    &    $\num{5e-04}^{+\num{1e-04}}_{-\num{1e-04}}$ &    \textcolor{red}{2.0*}   &    3.6       \\
       
       70059-03-01-01           &    52528.96521 & $800.0^{+0.2}_{-0.2}$       &    $114 \pm 42$ &    18.0      &    $1048^{+24}_{-24}$   &     \textcolor{red}{9.5*}   &    3.1       &    $41^{+4}_{-4}$    &    $\num{9e-04}^{+\num{2e-04}}_{-\num{2e-04}}$ &     \textcolor{red}{2.0*}   &    4.4       \\
       
       70059-03-02-00           &    52531.20519 &    $896^{+1}_{-1}$    &    $123 \pm 73$ &    13.2      &    $1106^{+11}_{-11}$   &     \textcolor{red}{21.1*}  &    3.7       &    $42^{+6}_{-16}$    &    $\num{4e-04}^{+\num{1e-04}}_{-\num{1e-04}}$ &     \textcolor{red}{2.1*}   &    2.7       \\
       
       70059-03-02-01           &    52531.48810 &    $1064^{+19}_{-14}$   &    $21 \pm 10$  & 3.0          &    $762.7^{+0.4}_{-0.5}$    &    $48 \pm 7$  &    18.6      &    $41^{+4}_{-8}$    &    $\num{1.3e-03}^{+\num{3e-04}}_{-\num{3e-04}}$ &  \textcolor{red}{2.0*}      &    4.8       \\
       
       70059-01-26-00           &    52546.86793 &    $663^{+3}_{-4}$    &    $29 \pm 17$  &    3.4       &    $966^{+11}_{-11}$    &     \textcolor{red}{17.9*}  &    3.4       &    $32^{+1}_{-1}$    &    $\num{4.9e-03}^{+\num{6e-04}}_{-\num{6e-04}}$ &     \textcolor{red}{2.1*}   &    8.2       \\
       
       80406-01-02-02           &    52715.71240 &    $908^{+4}_{-5}$    & $50 \pm 16$     &    6.0       &    $1145^{+9}_{-9}$   &    $17 \pm 6$  &    4.6       &    $37^{+4}_{-3}$    &    $\num{5e-04}^{+\num{1e-04}}_{-\num{1e-04}}$ &     \textcolor{red}{2.0*}   &    4.5       \\
       
       80406-01-03-03           &    52722.11726 &    $512^{+3}_{-4}$    &    $29 \pm 15$  &    3.4       &    $819^{+6}_{-6}$    &    $13 \pm 4$  &    5.3       &    $22^{+1}_{-1}$    &    $\num{4e-03}^{+\num{4e-04}}_{-\num{4e-04}}$ &     \textcolor{red}{1.9*}   &    9.5       \\
       
    \end{tabularx}
    \label{table:tab1}
    \renewcommand{\arraystretch}{1.0}
    \end{center}

\end{landscape}

\begin{landscape} 


	\captionof{table}[Triplets]{\textbf{Incomplete Triplets}: Observations in which only two of the three QPOs forming a complete triplet were present, and either the lower kHz QPO ($l$kHz QPO) or the upper kHz QPO ($u$kHz QPO) is missing. A red starred value for a Q factor means that the width of the QPO in question was fixed in order for it to have a sufficiently high Q factor or significance, or both. Assigned numbers for observations in the second column correspond to the numbers in Figure \ref{fig:hfqpoSplitOurs}.}
	\small%
	\noindent

	\begin{center}
    \renewcommand{\arraystretch}{1.5}
    \begin{tabularx}{1.24\textwidth}{lcccccccccccc}
    \toprule
    
    \multicolumn{13}{c}{\textbf{Incomplete Triplets}} \\ \toprule
    
    \textbf{Obs ID} & \# & \textbf{MJD} & \textbf{$l$kHz QPO (Hz)} & \textbf{Q factor} & \textbf{Sign.} & \textbf{$u$kHz QPO (Hz)} & \textbf{Q factor} & \textbf{Sign.} & \textbf{LF QPO (Hz)} & \textbf{Amplitude}   & \textbf{Q factor} & \textbf{Sign.} \\ \toprule

	70069-01-03-01  &  1  &  52548.68155  &  --  &  --  &  --  &  $833^{+11}_{-12}$  &  \textcolor{red}{17.5*}  &  3.4  &  $20.7^{+0.8}_{-0.7}$  &  $\num{2.9e-03}^{+\num{9e-04}}_{-\num{9e-04}}$  &  \textcolor{red}{6.6*}  &  3.3  \\
    
    70069-01-03-06  &  2  &  52554.87674  &  --  &  --  &  --  &  $796^{+11}_{-12}$  &  \textcolor{red}{14.3*}  &  3.6  &  $17.2^{+0.4}_{-0.3}$  &  $\num{2.4e-03}^{+\num{7e-04}}_{-\num{7e-04}}$  &  $9 \pm 3$  &  3.7  \\
    
    50052-01-24-00  &  3  &  51672.68155  &  --  &  --  &  --  &  $689^{+12}_{-13}$  &  \textcolor{red}{11.1*}  &  3.5  &  $21.4^{+0.4}_{-0.4}$  &  $\num{2.1e-03}^{+\num{8e-04}}_{-\num{7e-04}}$  &  $12 \pm 7$  &  3.0  \\
    
    60052-03-01-00  &  4  &  52233.87058  &  $493^{+13}_{-13}$  &  $2.3 \pm 0.3$  &  9.4  &  --  &  --  &  --  &  $28.7^{+0.8}_{-1.2}$  &  $\num{1.2e-03}^{+\num{6e-04}}_{-\num{4e-04}}$  &  $5 \pm 3$  &  3.1  \\
    
    95334-01-03-11  &  5  &  55270.37815  &  $834^{+3}_{-3}$  & $37 \pm 16$ &  4.8  &  --  &  --  &  --  &  $50^{+3}_{-2}$  &  $\num{1.7e-03}^{+\num{5e-04}}_{-\num{5e-04}}$  &  \textcolor{red}{5.7*}  &  3.2  \\
    
    70059-03-02-03  &  6  &  52531.55685  &  $780.8^{+0.4}_{-0.4}$  &  $88 \pm 28$  &  15.1  &  --  &  --  &  --  &  $51^{+4}_{-3}$  &  $\num{1.3e-03}^{+\num{4e-04}}_{-\num{4e-04}}$  &  $2 \pm 1$  &  3.7  \\
    
    95334-01-03-05  &  7  &  55272.59963  &  --  &  --  &  --  &  $729.2^{+0.4}_{-0.5}$  &  $51 \pm 11$  &  13.4  &  $25.6^{+0.7}_{-0.6}$  &  $\num{7e-04}^{+\num{2e-04}}_{-\num{2e-04}}$  &  \textcolor{red}{6.5*}  &  4.0  \\

    \end{tabularx}
    \label{table:tab2}
    \renewcommand{\arraystretch}{1.0}
    \end{center}


\end{landscape}

\twocolumn

\section{HBO QPOs as second harmonics}
\label{AppB}
\begin{figure}
\centering
\includegraphics[width=0.50\textwidth]{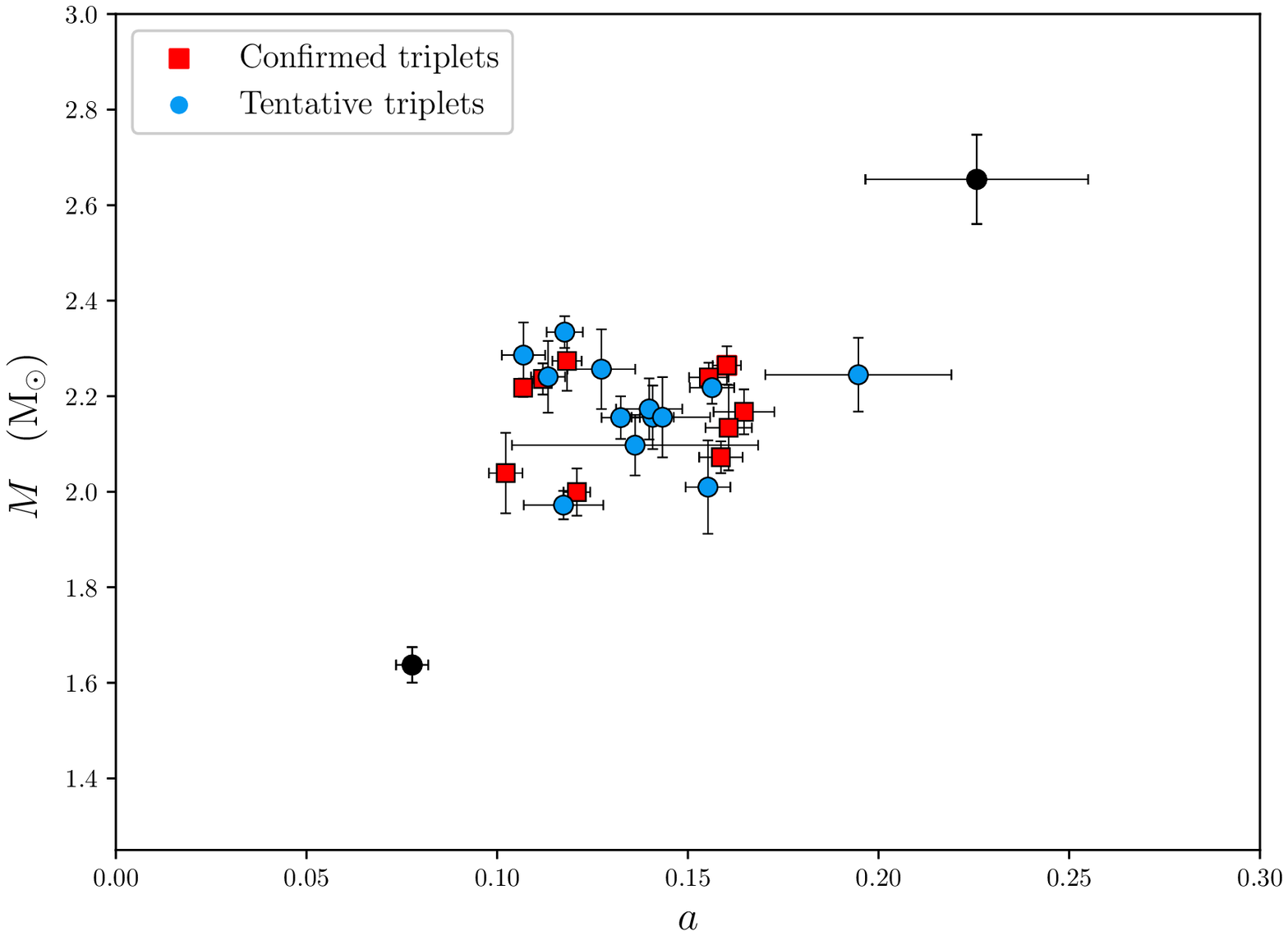}
\caption{The mass and spin estimates of the confirmed (red squares) and tentative (blue circles) triplets as predicted by the RPM, assuming that the measured HBO frequencies are in fact the second harmonics of non-detected fundamental frequencies. The black data points are the two tentative triplet outliers discussed in section \ref{subsec:rpmresults}.} 
\label{fig:appB}
\end{figure}

\cite{rpm2} noticed that the RPM yields high masses; as an explanation, they proposed that the HBO associated to a triplet could in fact be the second harmonic of the real Lense-Thirring frequency. Taking this into account and assuming that each of the HBOs we measured was the second harmonic of a non-detected fundamental frequency, we find the spin and mass ranges for 4U1608-52 to be $0.10 < a < 0.19$ and $1.97 < M/\textrm{M}_\odot < 2.33$, as can be seen in Figure \ref{fig:appB}.

\end{document}